\documentclass{pasj01}

\RequirePackage{mathpazo}
\RequirePackage[T1]{fontenc}

\newcounter{author}
\def\authorcount#1#2{\refstepcounter{author}\label{#1}
                     \altaffiltext{\ref{#1}}{#2}}

\begin{document}
\SetRunningHead{T. Kato}{Longest-period Dwarf Nova with Multiple Rebrightenings}

\Received{201X/XX/XX}%{yyyy/mm/dd}
\Accepted{201X/XX/XX}%{yyyy/mm/dd}

\title{ASASSN-14ho: Longest-period Dwarf Nova with Multiple Rebrightenings}

\author{Taichi~\textsc{Kato}\altaffilmark{\ref{affil:Kyoto}*}}

\authorcount{affil:Kyoto}{
     Department of Astronomy, Kyoto University, Kyoto 606-8502, Japan}
\email{$^*$tkato@kusastro.kyoto-u.ac.jp}

%%% end:list of authors


\maketitle

\begin{abstract}
The post-outburst rebrightening phenomenon in dwarf novae
and X-ray novae is still one of the most challenging subjects
for theories of accretion disks.  It has been widely recognized
that post-outburst rebrightenings are a key feature
of WZ Sge-type dwarf novae, which predominantly have
short ($\lesssim$0.06 d) orbital periods.
I found four post-outburst rebrightenings in ASASSN-14ho during
its 2014 outburst, whose orbital period has recently measured
to be exceptionally long [0.24315(10)~d].
Using the formal solution of the radial velocity
study in the literature, I discuss the possibility that this object
can be an SU UMa-type dwarf nova near the stability border
of the 3:1 resonance despite its exceptionally long orbital period.
Such objects are considered to be produced if mass transfer occurs
after the secondary has undergone significant
nuclear evolution and they may be hidden in
a significant number among dwarf novae showing multiple
post-outburst rebrightenings.
\end{abstract}

\section{Introduction}

   Cataclysmic variables (CVs) are close binaries consisting
of a white dwarf (primary) and a mass-transferring
red or brown dwarf (secondary).
The accreted matter around the primary forms an accretion
disk.  In some CVs, thermal instability in the accretion disk
causes outbursts and these CVs are called dwarf novae (DNe)
[see e.g. \citet{osa96review};
for general information of cataclysmic variables
and dwarf novae, see e.g. \citet{war95book}].
SU UMa-type DNe is a class of DNe showing long and bright
outbursts (superoutbursts) during which superhumps with
periods a few percent longer than
the orbital period ($P_{\rm orb}$) are present.
These superhumps are widely believed to be a consequence
of tidal instability resulting from the 3:1 resonance
in the accretion disk (\cite{whi88tidal}; \cite{osa89suuma};
\cite{hir90SHexcess}; \cite{lub91SHa}).
This 3:1 resonance is considered to be realized only
in systems with sufficiently low mass-ratios
($q=M_2/M_1$, where $M_1$ and $M_2$ are masses of
the primary and secondary, respectively).
WZ Sge-type DNe is an extreme extension of SU UMa-type DNe
with very small $q$, which enables to develop the 2:1 resonance
(\cite{lin79lowqdisk}; \cite{osa02wzsgehump})
[see \citet{kat15wzsge} for a modern review of
WZ Sge-type DNe].

   One of the defining characteristics of WZ Sge-type DNe is
the frequent presence of post-superoutburst
rebrightenings (cf. \cite{kat15wzsge}).  These post-superoutburst
rebrightenings are also called ``echo outbursts''
(cf. \cite{pat98egcnc}).  The origin of these post-superoutburst
rebrightenings is not still yet well understood.
Some authors claimed that they are a result of enhanced
mass-transfer from the secondary \citep{pat02wzsge},
while others proposed that they can be understood as
the increased turbulence immediately after a superoutburst
(\cite{osa01egcnc}; \cite{mey15suumareb}) even if
the mass-transfer is constant.
These has been a numerical simulation
which reproduced post-superoutburst rebrightenings by
artificially increasing the mass-transfer rate
\citep{bua02suumamodel}.  Although this model could reproduce
post-superoutburst rebrightenings, the disk shrank
in response to the inflow of matter of low specific
angular momentum, which is not supported by recent observations
of superhump periods in WZ Sge-type DNe (VSNET Collaboration,
in prep.).

   In the models of \citet{osa01egcnc} and \citet{mey15suumareb},
there is a need for a reservoir of mass in the accretion disk,
which enables multiple rebrightenings.  \citet{kat98super}
considered that a disk could expand beyond the radius of
the 3:1 resonance and that the matter beyond the resonance
could play a role of the reservoir.  This interpretation appears
to be confirmed by the strong infrared excess after
the superoutburst (e.g. \cite{uem08j1021}) [for a more complete
discussion, see \cite{kat15wzsge}].  This interpretation
is based on the assumption of an ``extra'' space beyond
the 3:1 resonance, which is only enabled in very low $q$,
corresponding to WZ Sge-type DNe.

   On the other hand, there have recently been a remarkable
advance of understandings of SU UMa-type near the stability border
of the 3:1 resonance.  A long-$P_{\rm orb}$ (0.09903~d)
dwarf nova V1006 Cyg showed a dip and
a post-superoutburst rebrightening \citep{kat16v1006cyg},
which had been only recorded in WZ Sge-type DNe.
CS~Ind ($P_{\rm orb}$=0.1242~d) showed a long precursor
outburst before a superoutburst, which had been also only
known in WZ Sge-type DNe \citep{kat19csind}.
These phenomena are considered to be a result of
the weak effect of the 3:1 resonance near the stability
border of the resonance, leading to premature quenching
of the superoutburst and resulting in a considerable
amount of matter after the superoutburst as is proposed
in WZ Sge-type DNe.

   Here I report multiple post-superoutburst rebrightenings
in a very long-$P_{\rm orb}$ (0.2432~d) in ASASSN-14ho.

\section{ASASSN-14ho}

   ASASSN-14ho was discovered by All-Sky Automated Survey for
Supernovae (ASAS-SN) Sky Patrol (\cite{ASASSN}; \cite{koc17ASASSNLC})
and confirmed as a DN by spectroscopy
\citep{pri14asassn14hoatel6619}.  The outburst started on
2014 September 9, and it did not catch special attention
since the phenomenon looked already over at the time of
notification in \citet{pri14asassn14hoatel6619}.
Although two rebrightening were already documented
in \citet{pri14asassn14hoatel6619},
the complete picture of the 2014 outburst had been unknown
before I obtained the full light curve from the ASAS-SN
archive on 2018 October 4 (vsnet-chat 8253).
In this vsnet-chat message,\footnote{
   VSNET chat message can be accessed at
   $<$http://ooruri.kusastro.kyoto-u.ac.jp/pipermail/vsnet-chat/$>$.
}
I wrote ``The object is too red for a WZ Sge-type dwarf nova.
Further study is recommended''.

   Quite recently, \citet{gas19asassn14hov1062cyg} reported
a radial velocity study of this object and obtained
$P_{\rm orb}$ of 0.24315(10)~d.  From the mass function,
\citet{gas19asassn14hov1062cyg} derived a constraint on
the masses of the primary, secondary and the inclination.
Their result favored a massive ($\sim$1.0 M$_{\odot}$)
white dwarf and an undermassive secondary for this $P_{\rm orb}$.

\section{Discussion}

\begin{figure*}
  \begin{center}
    \FigureFile(150mm,70mm){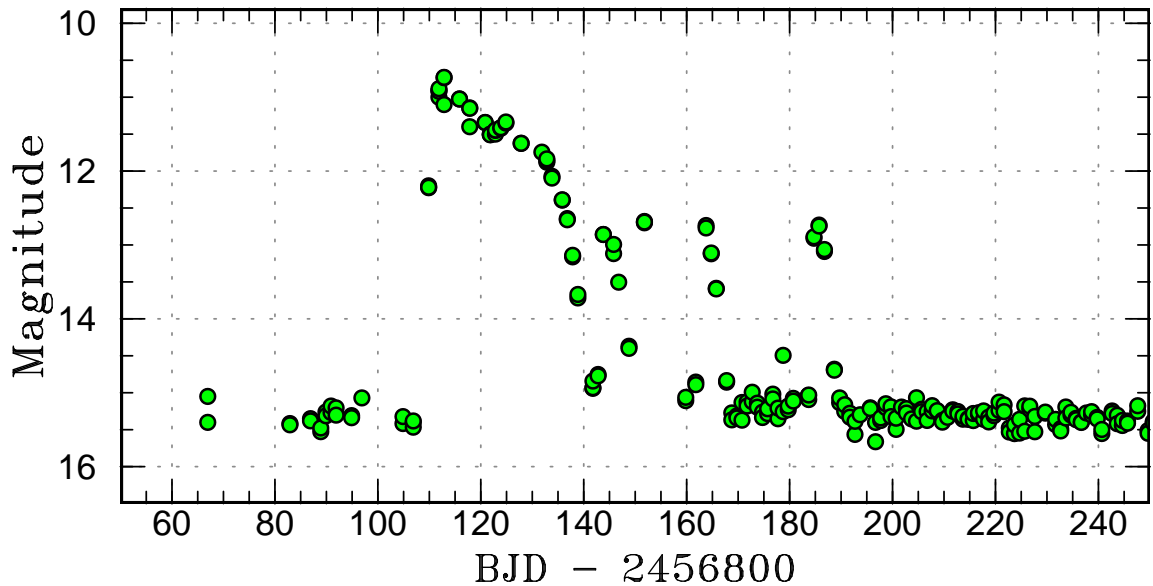}
  \end{center}
  \caption{Light curve of ASASSN-14ho during the 2014
  outburst from the ASAS-SN archive.  The main outburst
  was followed by four rebrightenings.
  }
  \label{fig:asassn14holc}
\end{figure*}

   The complete light curve of the 2014 outburst from
the ASAS-SN archive is shown in figure \ref{fig:asassn14holc}.
The main outburst was followed by four rebrightenings.
The $P_{\rm orb}$ of 0.24315(10)~d of ASASSN-14ho
is in the range of SS Cyg-type DNe (DNe without superhumps
or superoutbursts) [see a discussion in \citet{kat19csind}
for the borderline between SU UMa-type and SS Cyg-type DNe].
Such multiple rebrightening have been never observed
in SS Cyg-type DNe.

   Given the long $P_{\rm orb}$, ASASSN-14ho is extremely
unlikely a WZ Sge-type DN.  Could these rebrightenings
be caused by an enhanced mass-transfer in an SS Cyg-type
dwarf nova?  I consider that they can be interpreted
as a phenomenon seen in an SU UMa-type DN near
the stability border of the 3:1 resonance,
just like V1006 Cyg.  The most favorable masses in
\citet{gas19asassn14hov1062cyg} were $M_1$=1.0 M$_{\odot}$
and $M_2$=0.28 M$_{\odot}$, respectively, although they
were conservative enough to attempt a formal best fit.
If I accept them at face value, I obtain $q$=0.28,
which indeed close to the stability border of
the 3:1 resonance -- the modern values being 0.24
based on 3-D numerical simulation
\citep{smi07SH} or 0.33 under condition
of reduced mass-transfer \citep{mur00SHintermediateq}.
These value are also consistent with observations
of the suggested borderline objects
(\cite{kat16v1006cyg}; \cite{skl18nyser}).

   As for the $q$ value, this object appears to be
match perfectly the modern picture of an SU UMa-type
DN near the stability border of the 3:1 resonance
(confirmation, of course, requires direct detection of
superhumps during a future event).
Then the next question would be the consistency with
the long $P_{\rm orb}$.  \citet{gas19asassn14hov1062cyg}
rather conservatively stated this issue as
``the secondary of ASSASN-14ho is somewhat warmer and
significantly less massive than their fiducial sequence''.
It has been already observationally known such
a secondary is indeed present \citep{tho15asassn13cl}.
Although such a result was apparently not present in model
parameters in \citet{kni06CVsecondary} as stated in
\citet{gas19asassn14hov1062cyg}, studies have shown
that such systems can be formed if mass transfer occurs
after the secondary has undergone significant
nuclear evolution (e.g. \cite{pod03amcvn};
\cite{gol15CVevolution}).  In particular,
the $P_{\rm orb}$-$M_2$ diagram (figure 6) in 
\citet{gol15CVevolution} suggests that systems with
$P_{\rm orb}$=5.8 hr and $M_2$=0.28 M$_{\odot}$
(best values for ASASSN-14ho)
should be present as a significant population.
In \citet{mro16MCnova}, two objects were suggested to have
long $P_{\rm orb}$ and multiple rebrightenings
[OGLE-BLG-DN-0174 (0.14474(4) d) and OGLE-BLG-DN-0595 (0.0972(1) d)].
Although these periods were rather uncertain
[see Note added in proof in \citet{kat17j0026}],
long-$P_{\rm orb}$ DNe with multiple rebrightenings
may not be as rare as previously considered.
Further high-cadence wide-field observations
would unveil such a population.  I must, however, note
that final classification requires detection of superhumps
and high-cadence observations during the main outburst
(before multiple rebrightenings are confirmed) are essential.
This is still a challenge even for modern surveys.

\section*{Acknowledgments}

The author is particularly grateful to the ASAS-SN team for
making their data available to the public.

\end{document}